\documentclass[useAMS,usenatbib]{mnras}

\usepackage{epsfig,bm}
\usepackage{amsbsy,amssymb,amsmath}
\usepackage{graphicx}
\usepackage{dcolumn}
\usepackage[normalem]{ulem}

\usepackage{float}
\usepackage[usenames,dvipsnames]{xcolor}
\usepackage{hyperref}
\hypersetup{
    colorlinks = true,
    citecolor = {MidnightBlue},
    linkcolor = {BrickRed},
    urlcolor = {BrickRed}
}

\usepackage{ctable}
\usepackage{caption}

\newcommand{\uwc}{{Department of Physics \& Astronomy, University of the Western Cape, Cape Town 7535, South Africa}}
\newcommand{\qmu}{{School of Physics and Astronomy, Queen Mary University of London, Mile End Road, London E1 4NS, UK}}

\newcommand{\be}{\begin{equation}}
\newcommand{\ee}{\end{equation}}
\newcommand{\bea}{\begin{eqnarray}}
\newcommand{\eea}{\end{eqnarray}}

\title[Magnification measurements with HI intensity mapping]
{Prospects for cosmic magnification measurements using HI intensity mapping}

\author[]{Amadeus Witzemann$^{1}$\thanks{E-mail:amadeus.witzemann@gmx.at}, Alkistis Pourtsidou$^{2,1}$, Mario G. Santos$^{1, 3}$\\
$^{1}$\uwc\\
$^{2}$\qmu\\
$^{3}$South African Radio Astronomy Observatory (SARAO), 2 Fir Street, Observatory, Cape Town, 7925, South Africa
}

\date{Accepted XXX. Received YYY; in original form ZZZ}

\pubyear{2019}

\begin{document}

\maketitle

\begin{abstract}
We investigate the prospects of measuring the cosmic magnification effect by cross-correlating neutral hydrogen intensity mapping (HI IM) maps with background optical galaxies. We forecast the signal-to-noise ratio for HI IM data from SKA1-MID and HIRAX, combined with LSST photometric galaxy samples. We find that, thanks to their different resolutions, SKA1-MID and HIRAX are highly complementary in such an analysis. 
We predict that SKA1-MID can achieve a detection with a signal-to-noise ratio of $\sim 15$ on a multipole range of $\ell \lesssim 200$, while HIRAX can reach a signal-to-noise ratio of $\sim 50$ on $200 < \ell < 2000$.
 We conclude that measurements of the cosmic magnification signal will be possible on a wide redshift range with foreground HI intensity maps up to $z \lesssim 2$, while optimal results are obtained when $0.6 \lesssim z \lesssim 1.3$. Finally, we perform a signal to noise analysis that shows how these measurements can constrain the HI parameters across a wide redshift range.
\end{abstract}


\section{Introduction}
Traveling through the Universe, the path of light is deflected by the mass distribution it encounters. Images of distant light sources are distorted by the intervening matter along the line of sight (LOS), an effect well described by General Relativity. As a result, distortions of shapes, magnifications and even duplicate images are observed and are generally classified as weak or strong gravitational lensing. 

Weak gravitational lensing or cosmic shear is a coherent distortion of the shapes of galaxies, and has been routinely detected using optical galaxy surveys, with the first detections reported almost two decades ago (see, for example, \citet{Bacon:2000sy,Kaiser:2000if,vanWaerbeke:2000rm,Wittman:2000tc}). Ongoing and forthcoming large scale structure surveys like CFHTLens 
\citep{Heymans:2012gg}, DES \citep{Abbott:2016ktf}, Euclid \citep{Amendola:2016saw}, and LSST \citep{Abate:2012za}, will give precise cosmic shear measurements and use them to constrain the properties of dark energy. The accuracy and robustness of weak lensing measurements depends on the control of various systematic effects such as intrinsic alignments, point spread function, seeing and extinction, as well as photometric redshift calibration \citep{Mandelbaum:2017jpr}. In addition, Stage IV lensing surveys with Euclid and LSST will need accurate theoretical modelling of nonlinear clustering and baryonic effects down to very small scales to achieve their goals. Further improvements will come from the use of galaxy-galaxy lensing cross-correlations \citep{vanUitert:2017ieu}. 

In addition to the distortion of galaxy shapes, there is another form of lensing, cosmic magnification, which can be measured even when the sizes and shapes of sources are inaccessible. This makes it particularly attractive as it is free from many systematics such as the point spread function and intrinsic alignments (see, for example, \citet{2006MNRAS.367..169Z}, which discussed the possibility of using radio galaxy surveys to detect this effect). Magnification  occurs when intervening structure between an observer and a source acts to magnify or demagnify the object, i.e. sometimes allowing the observer to see objects otherwise too faint \citep{Bartelmann:1999yn}. However, the apparent observed area can also be increased, which leads to an apparent decrease in number counts if the total number is conserved. Only slightly altering the observed structures, this effect is notoriously difficult to measure (see e.g. the discussion in \citet{2009A&A...507..683H}). Several promising techniques exist, but there have been only a few, and controversial, detections (see discussion and references in \cite{2005ApJ...633..589S}). The first time this signal was measured with high significance was the $8\sigma$ detection achieved in \citet{2005ApJ...633..589S} using the Sloan Digital Sky Survey and the galaxy-quasar cross-correlation. A more recent analysis with DES galaxies is presented in \citet{Garcia-Fernandez:2016oud}.

Measurements of cosmic magnification probe the galaxy halo occupation distribution, dark matter halo ellipticities and the extent of galaxy dust halos \citep{2005ApJ...633..589S,Menard:2009yb} -- they are complementary to shear-shear measurements, and they can be used to break parameter degeneracies \citep{VanWaerbeke:2010yp}. 
Similar to cosmic shear, cosmic magnification provides constraints on the galaxy-matter correlation, but without the requirement of measuring shapes, it suffers from less systematic errors and can be extended to sources at much higher redshifts \citep{2005ApJ...633..589S}. In addition to probing the matter distribution directly, magnification also plays an important role in \emph{geometrical methods} to measure dark energy parameters independently of the matter power spectrum \citep{2003PhRvL..91n1302J,2004ApJ...600...17B,2007MNRAS.374.1377T}. These methods use galaxy-lensing correlations and therefore depend on estimates of the galaxy density. This is directly affected by magnification, which can therefore introduce systematic errors unless corrected for \citep{2005ApJ...633..589S,2007PhRvD..76j3502H,2008PhRvD..78l3517Z,Bonvin:2011bg}.

A straightforward approach to measure magnification uses the angular cross-correlation between foreground and background galaxy counts \citep[see e.g.][]{2009A&A...507..683H}, where galaxy-magnification or magnification-magnification cross-correlations would be major contributors to a non-zero signal. 

Following a similar line of thought, we propose to use HI intensity maps acting as foreground lenses, magnifying a background distribution of galaxies. A motivation for using HI is that intensity maps have no lensing corrections at first order due to flux conservation \citep{PhysRevD.87.064026}, which removes magnification-magnification correlations between foreground and background. This potentially decreases the signal, but also helps interpretation by removing additional terms in the signal calculation. In addition, the excellent redshift resolution of the foreground HI maps allows to combine measurements using different slices of the HI distribution. Using HI intensity maps also mitigates the danger of overlapping foreground and background sources, which results to a clustering (not lensing) signal. Furthermore, radio and optical observations are subject to different systematic effects, which are expected to drop out in cross-correlation. In the following, we derive forecasts for a potential detection of the magnification signal, using noise properties for the planned radio telescopes SKA1-MID \citep{Bacon:2018dui} and HIRAX \citep{2016SPIE.9906E..5XN}, as well as LSST.  

The plan of the paper is as follows: In section~\ref{sec:maggal} we give an introduction to cosmic magnification statistics and introduce the possibility of using HI intensity maps as foreground lenses. In section~\ref{sec:error} we calculate the instrumental (thermal) noise of SKA1-MID and HIRAX, as well as the shot noise from the LSST sample, and investigate the signal and noise properties for the cosmic magnification measurement. In subsection~\ref{ssec:original_results}, we optimise the signal-to-noise ratio for our proposed method and derive the cumulative signal-to-noise ratio for SKA1-MID and HIRAX. In subsection \ref{ssec:weighted_results}, we present constraints on the cumulative signal-to-noise ratio using a weighted galaxy over-density, which turns out to predict a better signal-to-noise. We also turn these constraints into fractional error forecasts on $\Omega_{\rm HI}b_{\rm HI}$. We summarise our findings and conclude in section~\ref{sec:conclusion}.

\section{Cosmic magnification statistics}
\label{sec:maggal}

In this section we describe the power spectrum formalism for measuring the cosmic magnification signal from background galaxies. We start with the standard approach, which assumes a galaxy sample as the foreground sample, and then introduce the possibility of using HI intensity maps instead.

\subsection{Galaxies as the foreground sample}

Galaxies are biased tracers of the underlying dark matter distribution, which is thought to contain most of the mass distributed along the LOS to a light source. Magnification will increase the flux from a galaxy, making it appear brighter than it actually is. Therefore galaxies normally too faint to be detected can still be seen if the magnification caused by the matter along the LOS is strong enough. However, the apparent area of a source is also increased, resulting in a decrease of the observed number density of galaxies. We can write \citep{2006MNRAS.367..169Z}
\be
\delta^{\rm L}_{\rm g}=\delta_{\rm g}+(5s_{\rm g}-2)\kappa +\mathcal{O}(\kappa^2) \, ,
\ee with $\delta_{\rm g}^{\rm L}$ and $\delta_{\rm g}$ the lensed and unlensed intrinsic galaxy over-densities, respectively, and $\kappa$ the lensing convergence. For a survey with limiting magnitude $m^\star$ the number count slope $s_{\rm g}$ is given by \citep{Duncan:2013haa}
\be
\label{eq:sg}
s_{\rm g} = \frac{d{\rm \, log}_{10}n_{\rm g}(< m^\star)}{dm^\star} \, ,
\ee
with the cumulative number of detected galaxies per redshift interval and unit solid angle, $n_{\rm g}$.
The cross-correlation of well separated foreground (at position $\theta_{\rm f}$ and redshift $z_{\rm f}$) and background ($\theta_{\rm b}$ and $z_{\rm b}$) galaxy samples is free from the intrinsic galaxy over-density correlation term $\langle \delta_{\rm g}(\theta_{\rm f},z_{\rm f}) \delta_{\rm g}(\theta_{\rm b},z_{\rm b}) \rangle$, therefore
\bea
\label{eq:magcorr}
\langle \delta^{\rm L}_{\rm g}(\theta_{\rm f},z_{\rm f}) \delta^{\rm L}_{\rm g}(\theta_{\rm b},z_{\rm b}) \rangle
&=& \langle(5s^{\rm b}_{\rm g}-2) \kappa_{\rm b}  \delta_{\rm g}(\theta_{\rm f},z_{\rm f}) \rangle \nonumber \\
&+& \langle(5s^{\rm f}_{\rm g}-2)(5s^{\rm b}_{\rm g}-2)  \kappa_{\rm f} \kappa_{\rm b} \rangle \, ,
\eea
where the superscript {\rm L} denotes lensed quantities. The right hand side of equation~\ref{eq:magcorr} contains the magnification-galaxy ($\mu-\rm g$) correlation (first term) and the magnification-magnification ($\mu-\mu$) correlation (second term). The latter is subdominant for foregrounds at comparably low redshifts and therefore usually neglected. If both foreground and background galaxies are at high redshifts, however, it can become large \citep{2008PhRvD..78l3517Z}.

\subsection{ HI intensity maps as the foreground sample}

In this work, we focus on the magnification effect of HI intensity maps in the foreground, acting on the clustering statistics of background galaxies. Intensity maps themselves are not lensed at linear order due to surface brightness conservation \citep{PhysRevD.87.064026}.  Intensity mapping lensing is very similar to CMB lensing, and a technique for measuring gravitational lensing in 21cm intensity mapping observations of HI after reionization was developed in \cite{Pourtsidou:2013hea}, building upon previous work by \cite{Zahn:2005ap}. In later studies, higher order effects were included. For example \citet{Jalivand:2018vfz} calculated the second order lensing effects on the intensity mapping power spectrum at $z=2-6$. They computed the corrections by Taylor-expanding in the deflection angle up to second order and found an extra term that can be important at high redshifts. 

To summarise, while the notion of number counts is not relevant for HI intensity mapping, the absence of lensing at linear order is formally equivalent to setting $s_{\rm HI} = 2/5$. We also have
\be
\delta T^{\rm L}_{21}=\delta T_{21}=\bar{T}_{21} \delta_{\rm HI} = \bar{T}_{21} b_{\rm HI} \delta \, ,
\ee where $\bar{T}_{21}$ is the mean brightness temperature of neutral hydrogen, $b_{\rm HI}$ is the hydrogen bias and $\delta$ the dark matter over-density. Considering galaxies as the background sample, we now have
\be
\langle \delta^{\rm L}_{\rm HI}(\theta_{\rm f},z_{\rm f}) \delta^{\rm L}_{\rm g}(\theta_b,z_b) \rangle
= \langle (5s^{\rm b}_{\rm g}-2) \kappa_b  b_{\rm HI} \delta (\theta_{\rm f},z_{\rm f}) \rangle \, ,
\ee
where the magnification-magnification term is absent since $s_{\rm HI}=2/5$. The above relation holds at all redshifts, given that the foreground and background samples are well separated. This can be guaranteed via the excellent redshift information provided by the intensity mapping survey.

The observable magnification signal can be expressed using the angular power spectrum \citep{2006MNRAS.367..169Z,2008PhRvD..78l3517Z}
\bea
\label{eq:clhimu}
&&C^{{\rm HI} - \mu}_\ell (z_{\rm f}, z_{\rm b}) = \frac{3}{2}\frac{H^2_0}{c^2}\Omega_{{\rm m},0} \times  \nonumber \\
&& \int^{\infty}_0 dz \frac{b_{\rm HI}\bar{T}_{21}(z)W(z,z_{\rm f})g(z,z_{\rm b})}{r^2(z)}(1+z) \times \nonumber \\
&& P_{\rm m}((\ell+1/2)/r(z),z) \, ,
\eea
where $r(z)$ is the comoving distance to redshift $z$ and we have applied the Limber approximation, valid for $\ell \geq 10$ \citep{1954ApJ...119..655L,2008PhRvD..78l3506L}. The redshift distribution of the foreground HI intensity maps is given by a top hat over the foreground redshift bin $W(z,z_{\rm f})$ and $g(z,z_{\rm b})$ is the lensing kernel:
\be
g(z,z_{\rm b})=\frac{r(z)}{N_{\rm g}(z_{\rm b})}\int^{z_{\rm b}^{\rm max}}_{z_{\rm b}^{\rm min}} dz' \frac{r(z')-r(z)}{r(z')}(5s_{\rm g}(z')-2)n_{\rm g}(z') \, ,
\label{eq:gkernel}
\ee
where the average number of galaxies per square degree in the background bin is
\be
N_{\rm g}(z_{\rm b}) \equiv \int_{z_{\rm b}^{\rm min}}^{z_{\rm b}^{\rm max}} n_{\rm g}(z)dz \, ,
\label{eq:ngNg}
\ee
and $z_{\rm b}^{\rm min}$, $z_{\rm b}^{\rm max}$ denote the minimum and maximum redshift for the background galaxy sample. An interesting feature of the geometrical weight $\frac{r(z')-r(z)}{r(z')}$ is that, in a flat universe, it takes the form of a parabola with a maximum at $r(z')=r/2$. Thus, structures half-way between the source and the observer are the most efficient to generate lensing distortions \citep{Kilbinger:2014cea} (and very low redshift foregrounds are less favoured).

\begin{figure}
\centering\includegraphics[width=0.95\columnwidth]{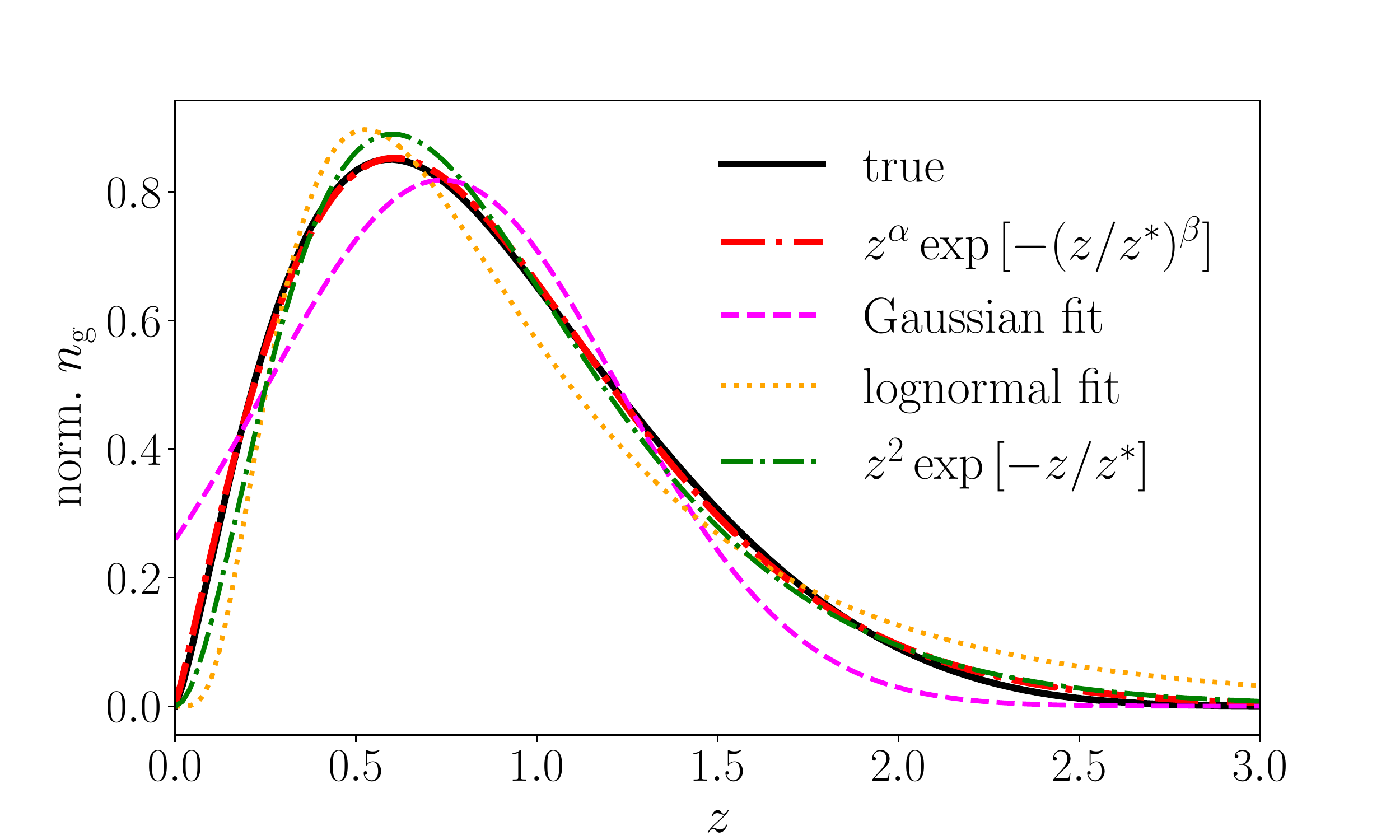}
\caption{Different fitting functions for the cumulative galaxy number count were considered. The normalised `true' function here is taken from \citet{0004-637X-814-2-145} (solid black line). The best fitting function (red dotted-dashed line) is given in equation~(\ref{eq:ngfit}).}
\captionsetup{width=.9\linewidth}
\label{fig:ngfits}
\end{figure}

\begin{figure}
\centering\includegraphics[width=0.95\columnwidth]{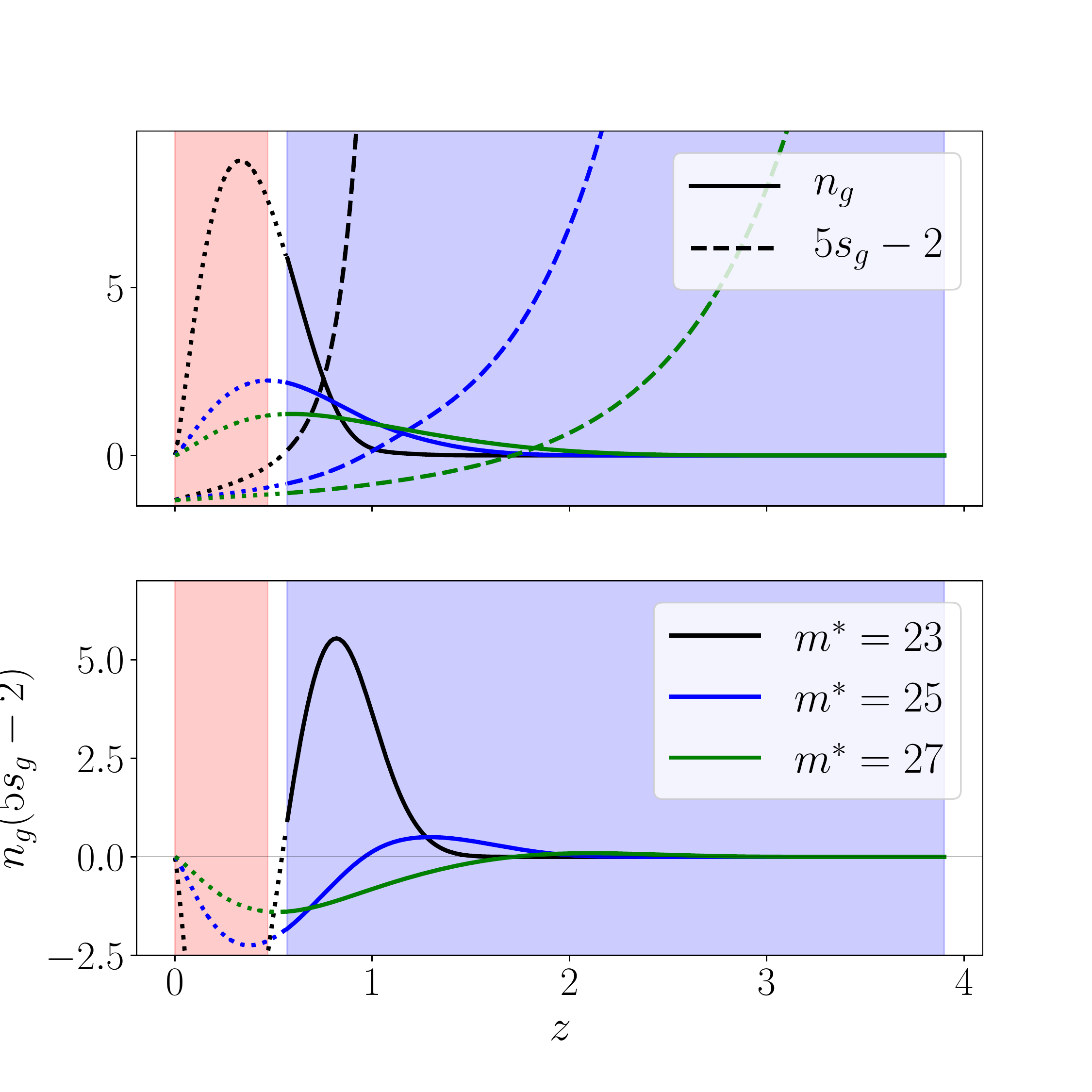}
\caption{We illustrate the behavior of the number count slope $s_{\rm g}$ and galaxy count $n_{\rm g}$ with respect to the magnitude threshold $m^*$, here with foreground redshift $0<z<0.47$, which corresponds to band 2 of SKA1-MID, described in detail in section \ref{sec:error}. The red (blue) shaded areas indicate the foreground (background) redshift range. The upper panel displays the galaxy number density $n_{\rm g}$ (normalised to integrate to one inside the background bin), and the contribution of the number count slope $s_{\rm g}$. The bottom panel shows the product $n_{\rm g}(5s_{\rm g} -2)$, which is the only term inside the integral Equation~(\ref{eq:gkernel}) to potentially be negative. This demagnification leads to cancellation in the integration and thus to a smaller lensing signal. An appropriate magnitude cutoff enforces $5s_{\rm g} >2$ in the background redshift bin and thus boosts the signal. However, this comes at the cost of increasing the galaxy shot noise.}
\label{fig:gkernel}
\end{figure}

\begin{figure}
\centering\includegraphics[width=.95\columnwidth]{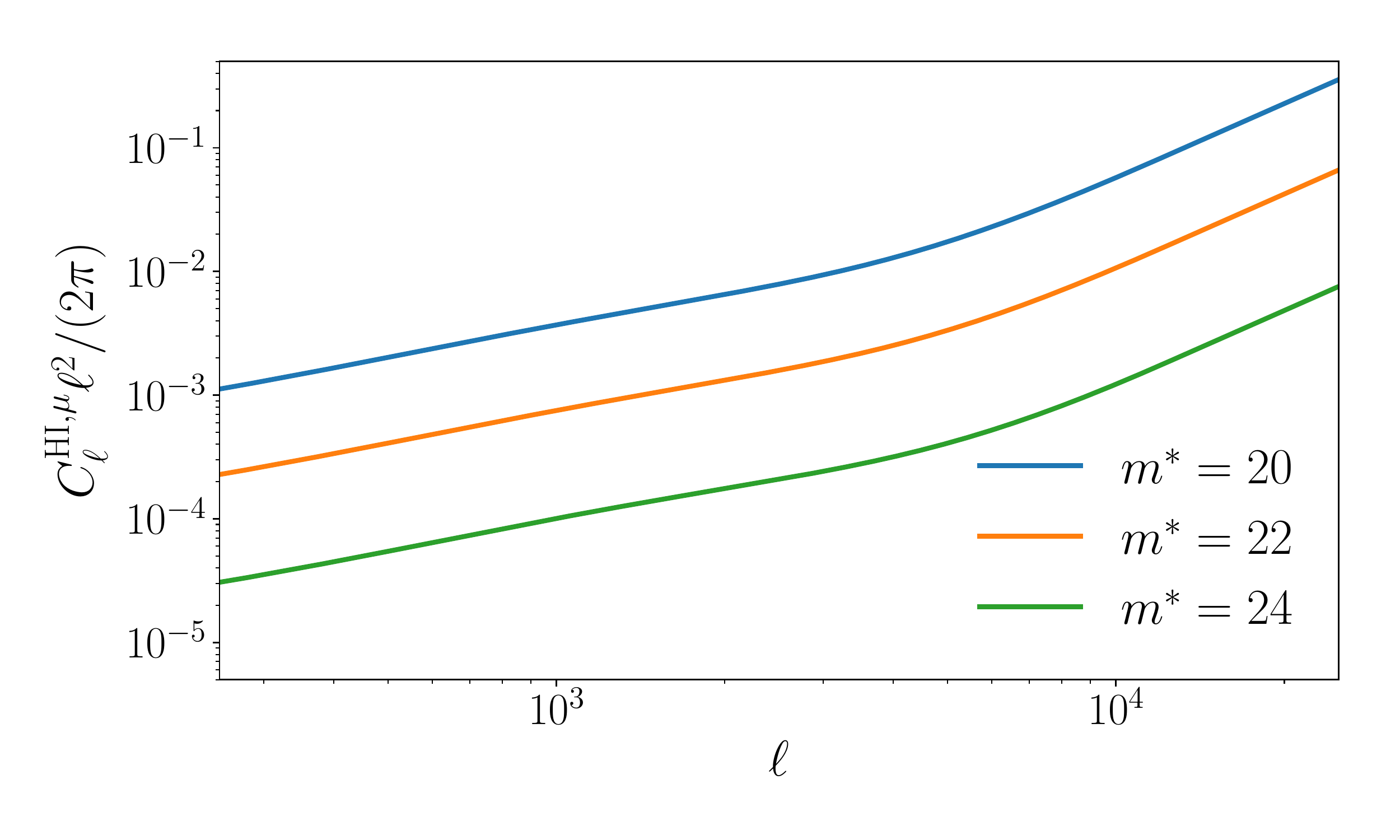}
\caption{The HI-magnification cross-correlation power spectrum for a foreground redshift from $z=0$ to $0.47$, corresponding to band 2 of SKA1-MID. Lower magnitude cuts increase the magnification signal.}
\label{fig:HIxmag_Cls}
\end{figure}

For increased computational speed, we use a fitting function to approximate the cumulative galaxy count for LSST, $n_{\rm g}$, provided in the publicly available code from \cite{0004-637X-814-2-145}. This code in turn uses the Schechter function \citep{1976ApJ...203..297S} for the r'-band luminosity from \cite{2006A&A...448..101G}, with the faint end slope $\alpha = -1.33$, the characteristic magnitude $M^*$

\be 
M_*(z) = M_0 + a \ln (1+z)
\ee
and the density $\phi^*$
\be
\phi_*(z) = (\phi_0 + \phi_1 z + \phi_2 z^2)[10^{-3} {\rm Mpc}^{-3}].
\ee
Here $M_0=-21.49$, $a=-1.25$, $\phi_0=2.59$, $\phi_1 = -0.136$, $\phi_2 = -0.081$.
We adapt the fit from \cite{2009arXiv0912.0201L} to approximate $n_{\rm g}$ as follows,
\be
n_{\rm g}(z) \propto z^\alpha \exp \bigl( -\bigl(\frac{z}{z^*}\bigr)^\beta \bigr),
\label{eq:ngfit}
\ee
where we optimise the parameters $\alpha$, $\beta$ and $z^*$ to fit $n_{\rm g}$ from \cite{0004-637X-814-2-145} as functions of magnitude cutoff $m^*$ by interpolation. Fig.~\ref{fig:ngfits} compares this fit with the true $n_{\rm g}$ and with several other fitting functions.  The overall amplitude is irrelevant in Equation~(\ref{eq:gkernel}), as $n_{\rm g}$ is normalised to integrate to one, but it is required to calculate the shot noise -- see section~\ref{sec:error} for details.

The number count slope $s_{\rm g}$ (Fig.~\ref{fig:gkernel}) rises quicker for a lower magnitude cutoff, therefore the magnitude threshold can be chosen to avoid a sign change of $5s_{\rm g} -2$ in the background redshift bin. The amplitude of the magnification signal is proportional to a redshift integral of $5s_g-2$ (Equations~(\ref{eq:sg}) and (\ref{eq:clhimu})). An appropriate magnitude cutoff thus boosts the signal by avoiding cancellations inside the integral for the lensing kernel $g$.  Fig.~\ref{fig:gkernel} demonstrates this in a situation where a lower magnitude threshold is beneficial to optimise the magnification signal, which is shown in Fig.~\ref{fig:HIxmag_Cls}. Decreasing $m^*$ comes at the cost of a smaller number of observed galaxies and therefore increased shot noise. We optimise to achieve a maximal signal to noise ratio. We will further discuss this in section~\ref{ssec:photo}, and we also note that a number count slope weighting was suggested in \citep{Menard:2002vz} and used in the SDSS data analysis of \citet{2005ApJ...633..589S}.

We use \texttt{CAMB} with \texttt{HALOFIT} \citep{2000ApJ...538..473L,Smith:2002dz,Takahashi:2012em} to estimate the nonlinear matter power spectrum, $P_{\rm m}(k,z)$, assuming a flat $\Lambda$CDM cosmology with $h=0.678$, $\Omega_ch^2=0.119$, $\Omega_bh^2=0.022$, $n_{\rm s}=0.968$ \citep{Aghanim:2018eyx}.

The error in the measurement of the cross-correlation power spectrum is
\be
\label{eq:magerr}
\Delta C^{{\rm HI} - \mu}_\ell = \sqrt{\frac{2((C^{{\rm HI} - \mu}_\ell)^2+(C^{\rm gg}_\ell+C^{\rm shot})(C^{\rm HI-HI}_\ell+N_\ell))}{(2\ell+1)\Delta \ell f_{\rm sky}}} \, ,
\ee where $C^{\rm shot}$ is the galaxy shot noise power spectrum, $N_{\rm \ell}$ is the thermal noise of the intensity mapping instrument, $\Delta \ell$ is the binning in multipole space, and $f_{\rm sky}$ is the fraction of sky area overlap of the HI and optical surveys.  

\begin{figure}
\centering\includegraphics[width=0.95\columnwidth]{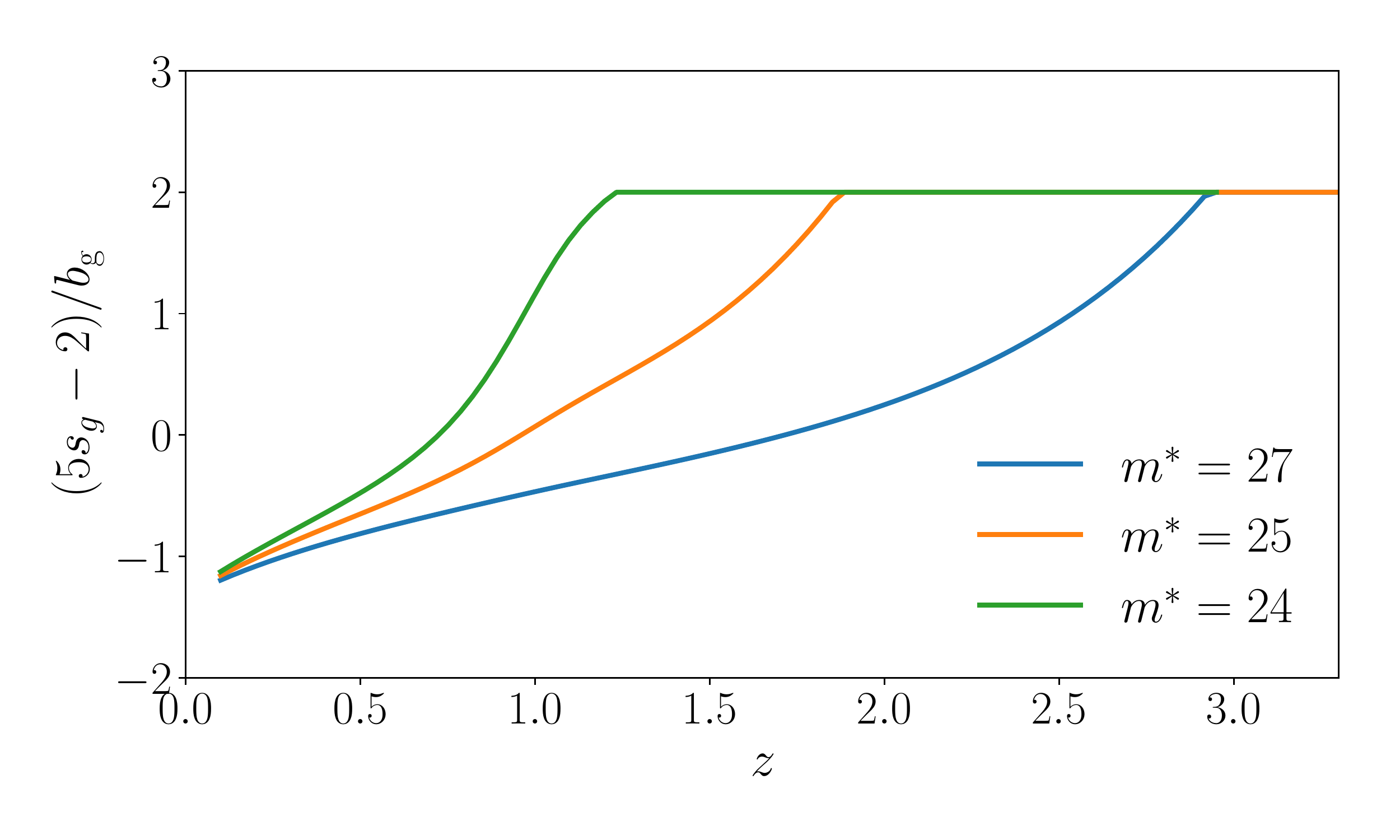}
\caption{The ratio of number count slope and galaxy bias $(5s_{\rm g}-2)/b_{\rm g}$ for different magnitude cutoff values. This sample dependent ratio describes the relative importance of clustering and magnification.}
It is set to $2$ for higher redshifts via the choice of galaxy bias, see Equation~(\ref{eq:mag_gbias}). This choice has a negligible effect on our results, as it affects a regime where the number of galaxies quickly approaches zero.
The maximum magnitude detectable with LSST is assumed to be $27$.
Imposing a lower magnitude cutoff increases the shot noise, but also the number count slope, which increases the magnification signal. 
\label{fig:sg}
\end{figure}

For the foreground HI IM sample we use a top-hat window function $W(z)=1/\Delta z$ inside the bin of width $\Delta z$ and zero elsewhere. We can then write the HI and galaxies auto-correlation power spectra as 
\begin{equation}
C_\ell^{\rm HI-HI} = \frac{H_0}{c} \int dz E(z) \biggl( \frac{ b_{\rm HI}\bar{T}_{21}(z) W(z)}{r}\biggr)^2 P_{\rm m} \biggl( \frac{\ell + 1/2}{r}, z \biggr),
\end{equation}
and 
\begin{equation}
  \label{eq:clgal}
C_\ell^{{\rm g-g}} = \frac{H_0}{cN_{\rm g}^2} \int dz E(z) \biggl( \frac{ b_{\rm g }(z) n_{\rm g}(z)}{r}\biggr)^2 P_{\rm m} \biggl( \frac{\ell + 1/2}{r}, z \biggr),
\end{equation}
where we have written the Hubble rate as $H(z) = H_0E(z)$, and the HI bias $b_{\rm HI}$ is given by fits to the results from \cite{0004-637X-814-2-145}:
\begin{equation}
  b_{\rm HI}(z) = 0.67 + 0.18 z + 0.05 z^2 \, .
\end{equation}
The galaxy bias $b_{\rm g}$ naturally depends on redshift as well as magnitude cutoff, as brighter objects are rarer and thus more biased, an effect which is ignored when a simple linear and dererministic fitting function is used, for example
\begin{equation}
  \tilde{b}_{\rm g}(z) = 1+0.84 z.
  \label{eq:mag_gbiasfit}
\end{equation}
To enforce a behaviour similar to that of the magnification bias at higher redshifts and more stringent magnitude cuts, we use a piecewise differentiable galaxy bias:
\begin{equation}
  b_{\rm g}(z) = {\rm max}\biggl( \tilde{b}_{\rm g}, \frac{1}{2} (5s_{\rm g}-2)\biggr) \, .
  \label{eq:mag_gbias}
\end{equation}
This choice makes sure that the ratio $(5s_{\rm g}-2)/b_{\rm g}$ converges for $n_\mathrm{g} \to 0$, as described in \citep{2007PhRvD..76j3502H}.
The value of this ratio depends on the selected sample, and it describes the relative importance of clustering and magnification. Measurements of the redshift dependence of the galaxy luminosity function can be used to derive constraints using the halo model \citep[see e.g.][]{2007PhRvD..76j3502H,2007PhRvD..75d3519L}.
In our forecasts, this ratio is shown in  Fig.~\ref{fig:sg} for different magnitude cutoff values. As already mentioned, we are  setting it to $2$ for higher redshifts to avoid computational issues in a regime which is already heavily shot-noise dominated \citep{2007PhRvD..76j3502H}. We emphasise that while this choice is an assumption, it is a justified one -- with a negligible effect on our analysis. The signal remains unaltered and (as will be shown) errors are mostly shot noise dominated.

The mean observed HI brightness temperature is calculated using the fit provided in \cite{2017arXiv170906099S}, which is based on the results from \cite{2015aska.confE..19S}:
\be
\bar{T}_{21} = 0.0559+0.2324z-0.024 z^2 {\rm~mK} \, .
\ee

\section{Error calculations}
\label{sec:error}

\subsection{HI intensity maps}

We consider the experiments HIRAX and SKA1-MID to map the distribution of HI, used as the foreground sample. Together with the shot noise from LSST, their instrumental noise contributes to the total error budget given by Equation~(\ref{eq:magerr}).

HIRAX is a planned radio interferometer of $6$ m diameter dishes, sharing the site in the Karoo in South Africa with MeerKAT and SKA1-MID. We assume the full planned array of 1024 and the reduced set of 512 dishes, arranged in a dense square grid with $1$ m space between individual antennas. HIRAX aims to perform a large sky intensity mapping survey with $15,000$ deg${}^2$ area, and the integration time is taken to be two full years (corresponding to 4 years observation). We assume a constant system temperature of 50 K on its entire frequency coverage ranging from $400$ to $800$ MHz \citep{2016SPIE.9906E..5XN}.

At the same time, SKA1 is assumed to have only one year worth of integration time but a larger survey area of 16,900 deg${}^2$. This corresponds to the maximum possible survey overlap with LSST, after taking into account the total survey area of SKA1-MID \citep{2015aska.confE..19S} and contamination from galactic synchrotron radiation and dust. SKA1-MID will consist of different dish types: the (already operating) 64 MeerKAT dishes with $13.5$ m, and $133$ SKA1-MID dishes of $15$ m diameter. For simplicity, we assume all dishes to be identical, taking an average dish diameter $\tilde{D}_\mathrm{dish} = (64 \times 13.5 + 133 \times 15)/(64+133){\rm~m}$ and using a Gaussian beam pattern. We consider two observational bands: band 1 ranging from $350$ to $1050$ MHz, and band 2 from $950$ to $1750$ MHz \citep{Bacon:2018dui}. The system temperature is assumed to be  $30$ K for band 1 and $20$ K for band 2. This is conservative on low redshifts. For high-redshift foreground bins, the system temperature increases beyond that, but at the same time the galaxy shot noise becomes the dominant source of error and magnification detections quickly become extremely difficult for foreground samples with $z \gtrsim 2$. This justifies our assumption of constant system temperature for both SKA1-MID and HIRAX. For both experiments, we use equally spaced redshift bins of width $\Delta z = 0.5$, with the exception of band 2 with $\Delta z = 0.47$. A more realistic treatment would have to take into account the frequency dependence of the noise temperatures of both experiments, and the different dish and receiver types of SKA1. However, we expect this to have a negligible effect on our results.

Following \cite{Battye:2012tg} and \cite{0004-637X-803-1-21} for the intensity mapping noise calculations, we calculate the single dish noise for SKA1-MID as
\begin{equation}
  N^{\rm SD}_\ell = \sigma_\mathrm{pix}^2\Omega_\mathrm{pix}W_\ell^{-1} \, .
\end{equation}
Here, we use the solid angle per pixel $\Omega_\mathrm{pix} = 4\pi f_{\rm sky}/N_\mathrm{pix}$, the number of pixels $N_\mathrm{pix}$, the beam ($\Theta_\mathrm{FWHM}$) smoothing function $W_\ell = \exp\bigl(-\ell^2 \Theta_\mathrm{FWHM}^2/(8\ln{2})\bigr)$, the pixel noise $\sigma_\mathrm{pix} = T_\mathrm{sys}  \sqrt{ N_\mathrm{pix} / ( t_\mathrm{tot} \delta_\nu N_\mathrm{dish})}$ and the frequency resolution (channel width) $\delta_\nu$.

For HIRAX, we calculate the interferometer noise
\begin{equation}
  N^{\rm INT}_\ell = \frac{(\lambda^2 T_{\rm sys})^2}{2 A_{\rm e}^2 d\nu n(u) t_{\rm p}} \, ,
\end{equation}
with the frequency bin width $d\nu$, the time per pointing $t_{\rm p} = t_{\rm tot} / N_{\rm p}$, the effective collecting area of one dish $A_{\rm e} = (D_{\rm dish}/2)^2 \pi$, and using the relation $u = \ell / (2 \pi)$ for the baseline density $n(u)$.

For all experiments we assume full survey overlap with LSST.
\subsection{Photometric galaxy counts}
\label{ssec:photo}

We normalise the LSST sample to be a total of $\sim 6.3\times 10^9$ galaxies at $m^*=27$\footnote{This is slightly more conservative than the number quoted in \citet{2009arXiv0912.0201L}, i.e. almost $10^{10}$ galaxies for $m^*=27.5$.}.
The galaxy shot noise for LSST is calculated as $C^\mathrm{shot} = 4\pi/N_{\rm g}^{\rm LSST}(z)$, where we use a fitting function to calculate the number of detected galaxies in the considered redshift bin, $N_{\rm g}^{\rm LSST}$ (Eqs.~(\ref{eq:ngNg}) and (\ref{eq:ngfit})). We consider all possible LSST redshift bins to have their upper edge at the same $z_{\rm max}^{\rm LSST} = 3.9$, and the lower bin edge at a separation from the upper edge of the foreground bin, $z^{\rm fg}_i + 0.1$. 
The choice of a separation of $\Delta z = 0.1$ might not completely rule out cross-correlations (because of photometric redshift outliers), but should be enough to reduce them to a very low level. We calculate the number count slope for LSST using an adjusted version of the code provided in \cite{0004-637X-814-2-145} to extend to more stringent luminosity cutoffs $m^*$. We then interpolate $(5s_{\rm g}-2)n_{\rm g}$ on a fine grid ($z$ and $m^*$) to speed up the numerical calculations.

\begin{figure}

\centering\includegraphics[width=.99\columnwidth]{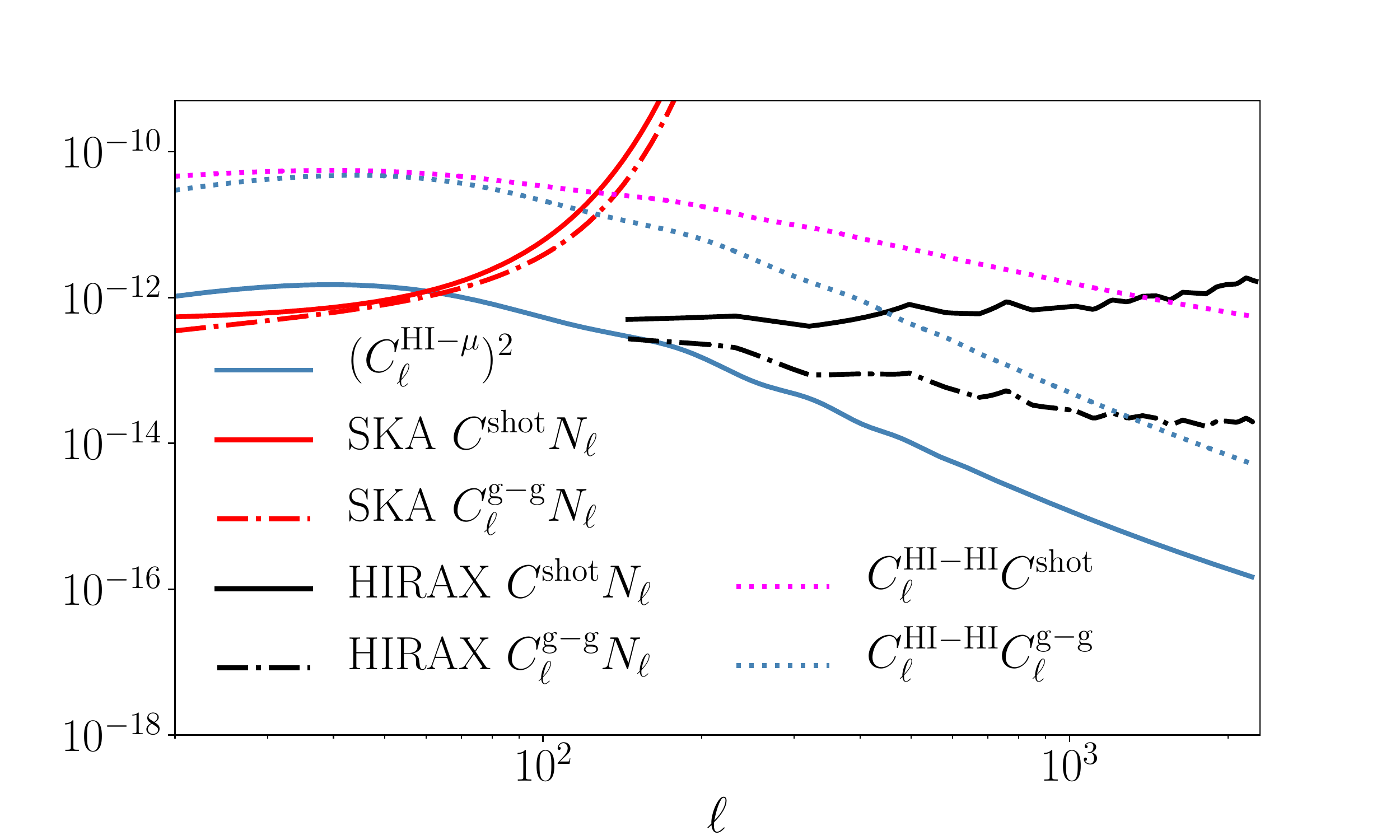}
\captionsetup{width=.9\linewidth}

\caption{All contributions to $\Delta C_\ell^{\mathrm{HI-}\mu}$ as in Equation~(\ref{eq:deltah-mu}), for a foreground redshift bin from $z=0.85$ to $1.35$ and a background bin $z\geq1.45$. The common factor of $2/((2\ell+1)\Delta\ell f_{\rm sky}$ was omitted here. Terms proportional to the SKA1-MID (HIRAX) noise are plotted in red (black) and terms proportional to shot noise and cosmic variance are plotted cyan and steel blue respectively. For this choice of binning, the HI intensity mapping noise is subdominant, followed by pure cosmic variance, but both dominated by terms with shot noise, the biggest source of error. Note that the choice of intermediate foreground and background redshift in this plot is only to ease comparison, but not necessarily ideal for magnification measurements.}
\label{fig:x_error_contributions}
\end{figure}

In order to illustrate the different error contributions and consolidate our findings, Fig.~\ref{fig:x_error_contributions} shows all summands contributing to the HI-magnification cross correlation error:
\begin{align}
  (\Delta C_\ell^{{\rm HI-}\mu})^2 = \frac{2}{ (2\ell+1)\Delta\ell f_{\rm sky}} \biggl(   (C_\ell^{{\rm HI-}\mu})^2  + C_\ell^{\rm g-g}C_\ell^{\rm HI-HI}  \nonumber \\
  +  C^{\rm shot}C_\ell^{\rm HI-HI} + N_\ell C_\ell^{\rm g-g} +  C^{\rm shot}N_\ell \biggr).
  \label{eq:deltah-mu}
\end{align}
The amplitude of the different contributions here depends on the choice of experiments and redshift binning. 

To ease comparison we used the same single redshift bin for HIRAX and SKA1-MID in Fig.~\ref{fig:x_error_contributions}, from $z=0.85$ to $1.35$. For HIRAX a magnitude cutoff of $m^* = 24.4$ maximises the signal-to-noise ratio; for SKA1-MID it is $24.3$. This optimisation will be discussed further in section \ref{ssec:original_results}.
In this case shot noise dominates the error throughout, but it becomes comparable to cosmic variance (mostly $C_\ell^{\rm g-g}C_\ell^{\rm HI-HI}$) on large scales for SKA1-MID. Note that small scales are practically inaccessible for SKA1-MID due to its poor angular resolution, restricting it to much larger scales than HIRAX.

The multipole resolution is set by the maximum scale accessible by the SKA, i.e. the survey area $S_{\rm area}$ when in single dish mode. We estimate $\ell_{\rm min}^{\rm SKA} = 2\pi/\sqrt{S_{\rm area}} \sim 3$, but choose a more conservative value of $\ell_{\rm min}^{\rm SKA} = 10$ for the Limber approximation to hold \citep{2008PhRvD..78l3506L}. For the HIRAX interferometer it is set by the field of view ($\rm fov$) which depends on frequency. For the sake of simplicity we ignore this dependence and assume a mean ${\rm fov} = 35.5 {\rm~deg}^2$ \citep{2016SPIE.9906E..5XN}, giving $\ell_{\rm min}^{\rm HIRAX} = 2\pi/\sqrt{\rm fov} \sim 60$. From the signal to noise ratio $C_\ell^{{\rm HI} - \mu} / \Delta C^{{\rm HI} - \mu}_\ell$ we calculate the cumulative (total) signal to noise as

\be
{\rm SN}_{\rm tot} = \sqrt{\sum_{\bar{\ell}=\ell_{\rm min}}^\ell (C_{\bar{\ell}}^{{\rm HI} - \mu} / \Delta C^{{\rm HI} - \mu}_{\bar{\ell}})^2},
\label{eq:s2ncum}
\ee
where the sum runs over the relevant $\ell$ values, with the minimum $\ell$, and the binning $\Delta \ell$, set by $\ell_{\rm min}$. We note, however, that the cumulative signal to noise ratio ${\rm SN}_{\rm tot}$ is binning independent.

\section{Results and Discussion}
\subsection{Signal to noise ratio}
\label{ssec:original_results}

\begin{figure}
\centering\includegraphics[width=.99\columnwidth]{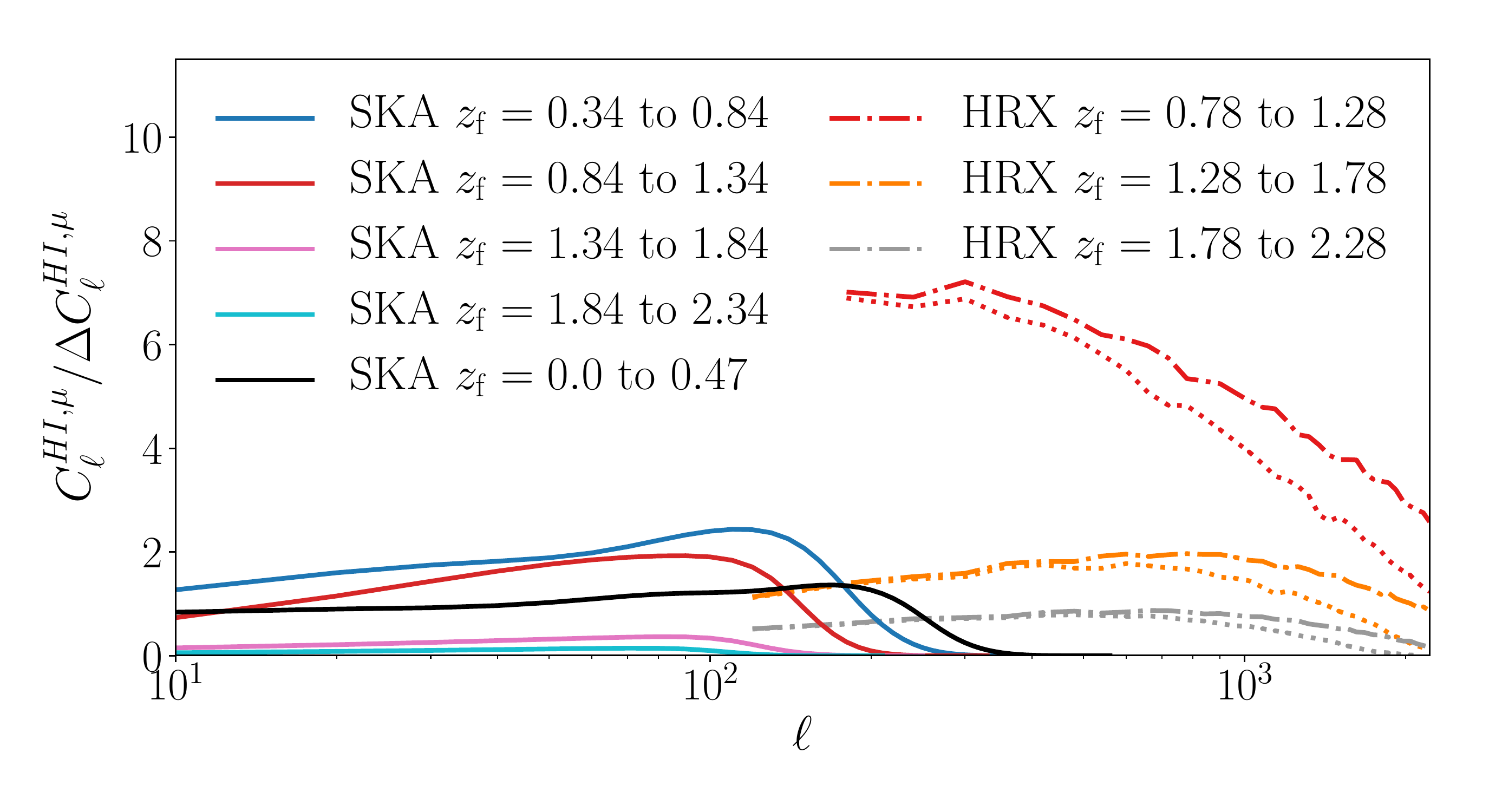}
\centering\includegraphics[width=0.99\columnwidth]{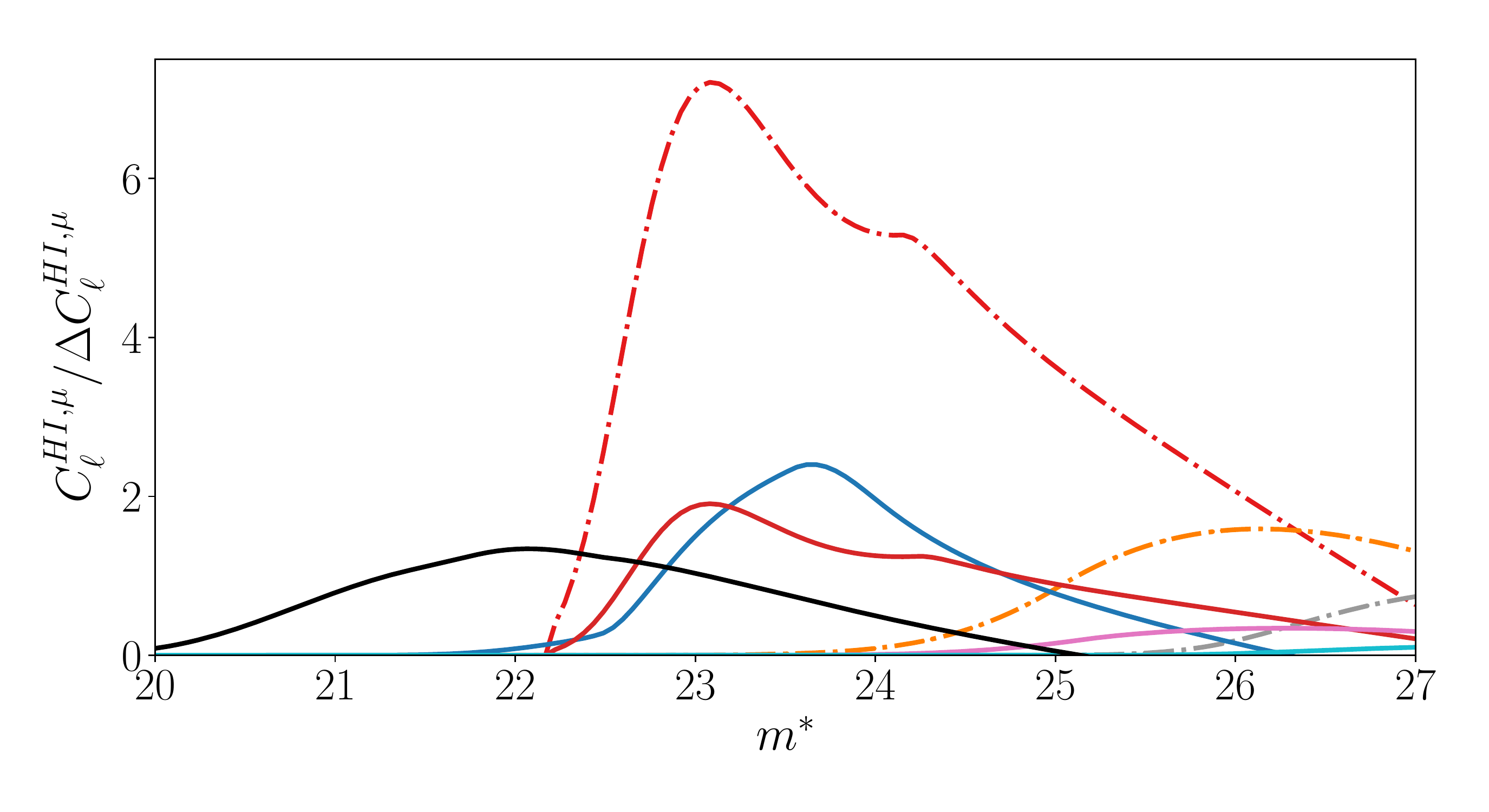}
\captionsetup{width=0.95\linewidth}
\caption{{\sl Upper panel:} The expected signal to noise ratio of the magnification signal for the combinations HIRAX 1024 (dotted-dashed lines), HIRAX 512 (dotted lines) and SKA1 band 1 (solid lines) and band 2 (solid black line) with LSST. We use different foreground redshift bins, always combined with one single non-overlapping background bin. Shot noise largely dominates, therefore the $512$ dish version of HIRAX performs surprisingly well compared to the full array with $1024$ dishes.
{\sl Lower panel:} The optimisation of the signal to noise ratio as a function of magnitude cutoff $m^*$. This panel is for single $\ell$ bins only, for SKA1-MID $\ell = 80$ and for HIRAX $\ell = 200$. These values were chosen to lie within the experiment's range of maximum sensitivity.}
\captionsetup{width=.9\linewidth}
\label{fig:S2N}
\label{fig:S2Nmstar}
\end{figure}

\begin{figure}
\centering\includegraphics[width=.99\columnwidth]{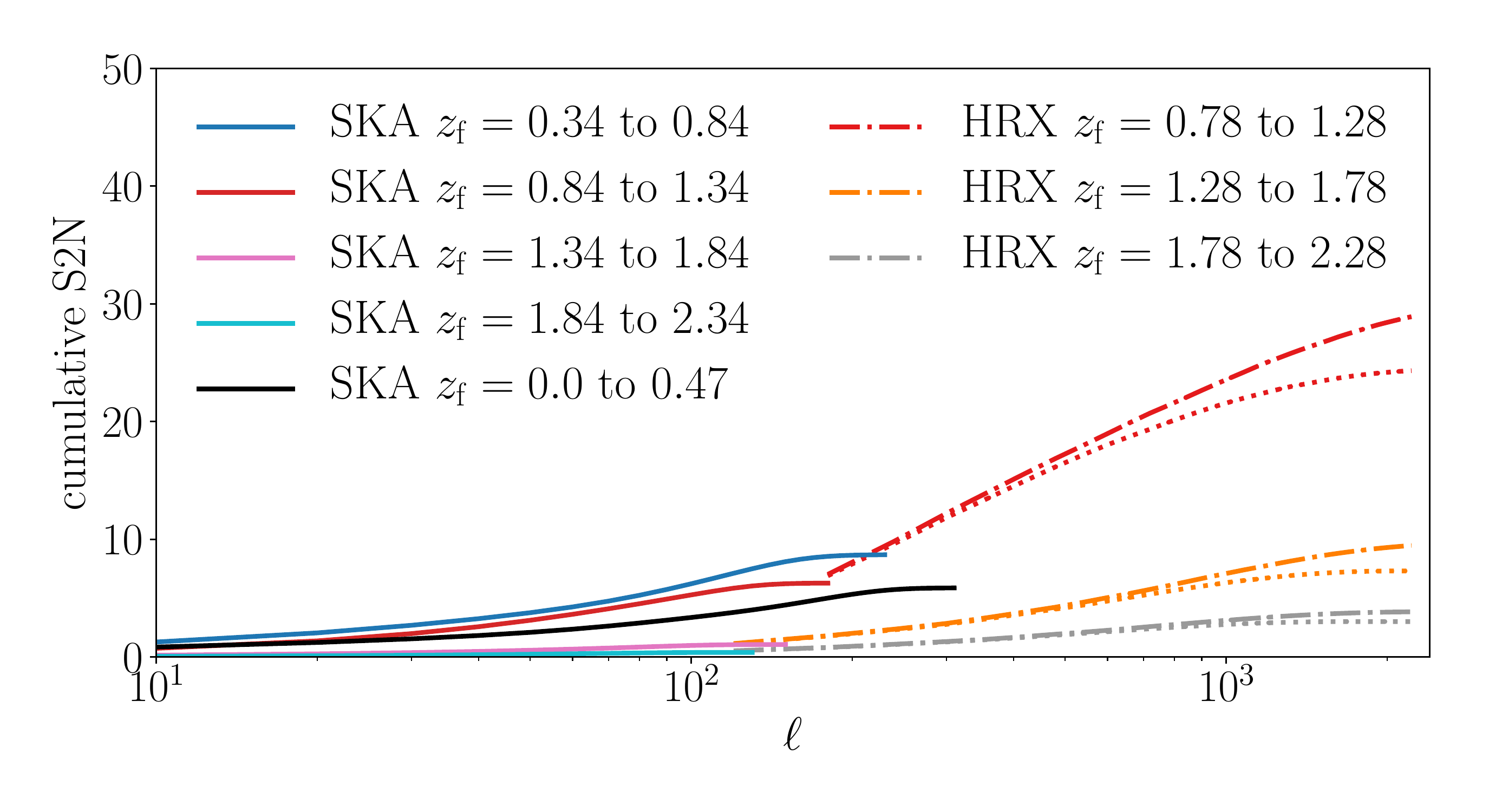}
\captionsetup{width=.9\linewidth}

\caption{This plot shows the cumulative signal to noise ratio, Equation~(\ref{eq:s2ncum}). For HIRAX especially, the error is dominated by the galaxy shot noise. Therefore even the down-scaled design with $512$ dishes yields very similar results compared to the full proposal with $1024$ dishes.}
\label{fig:S2N_cum}
\end{figure}

\begin{table}
  \begin{center}
    \begin{tabular}{l|cccc}
      &SKA1 B1&-&-&- \\
      $z$ range & 0.34-0.84 & 0.84-1.34 & 1.34-1.84 & 1.84-2.34 \\
      $m^*$           &  23.6 &  23.1 & 26.3 &  27.0 \\
      SN$_{\rm tot}$       &  8.7 &  6.3 & 1.1 &  0.4 \\
      \hline
       & HIRAX  & - & - & SKA1 B2\\
     $z$ range  & 0.78-1.28 & 1.28-1.78 & 1.78-2.28 & 0.0-0.47\\
     $m^*$            &  23.0 & 26.1 &  27.0&  22.1 \\
     SN$_{\rm tot}$        &  28.5 & 9.4 &  3.8 &  5.8\\
    \end{tabular}
    \captionsetup{width=.9\linewidth}

    \caption{Optimised magnitude cutoffs, $m^*$, as well as cumulative signal to noise values for all experiments and redshift bins. Individual redshift bins of HIRAX are better than SKA1-MID also due to the higher number of $\ell$ bins that contribute. }
    \label{tab:res}
  \end{center}
\end{table}

We maximise the signal to noise ratio with respect to the galaxy magnitude threshold $m^*$ for each HI survey and redshift bin. We consider an optimisation range of $m^* \in [19,27]$ and plot $\mathrm{SN}(m^*)_{\rm tot}$ for a few examples in Fig.~\ref{fig:S2Nmstar}. The optimal values we found (using the \emph{python} package \emph{scipy optimize}) are shown in Table~\ref{tab:res}. Generally, for low-redshift foreground bins, also a low $m^*$ is preferred, which increases the number count slope at the acceptable cost of increasing the (negligible) shot-noise at these redshifts. For high-redshift foreground bins, however, shot-noise increases and $m^*$ needs to be higher to account for this.

Fig.~\ref{fig:S2N} shows the optimised signal to noise as a function of multipole for all considered experiment and redshift combinations. Maps in each foreground redshift bin are correlated with one single redshift bin of LSST, separated from the foreground by $\Delta z =0.1$ and ranging up to $z=3.9$. Low redshift foreground bins benefit from a wider background sample containing a larger number of galaxies. Therefore, they often perform better than high redshift bins, especially in the case for HIRAX. The sensitivity of HIRAX is best at comparably small scales, where the power spectrum drops $\sim \ell^2$ (see e.g. Fig.~\ref{fig:HIxmag_Cls}). The shot noise, however, becomes the dominant error on smaller scales. The 512 dish design for HIRAX performs surprisingly well, as even in this case the interferometer noise remains subdominant.

Figure \ref{fig:S2N_cum} shows the cumulative signal to noise which reaches levels of $\sim 30$ for individual redshift bins. 
The performance of SKA1-MID and HIRAX is similar for single $\ell$ bins, but HIRAX covers a larger multipole range. Both experiments yield best results at intermediate redshifts of $0.6< z < 1.3$. As they are sensitive to different angular scales, most of their constraining power can be combined.

\subsection{Weighting analysis}
\label{ssec:weighted_results}
In this section we will investigate whether it is possible to further increase the signal to noise ratio by using the weighting proposed by \cite{Bartelmann:1999yn}. Let us suppose that galaxies are split into magnitude bins. We can then consider a weighted galaxy over-density $\sum_i \mathcal{W}_i \delta_{{\rm g},i}$, where summation runs over galaxy magnitude bins 
and $\mathcal{W}_i$ denotes a scale independent weighting function.

In Fourier space, we can derive the weighted signal to noise ratio, $({\rm SN}_\mathcal{W})$, adapting the formula derived in \citet{2011MNRAS.415L..45Y} for a 21cm 
intensity mapping foreground sample. We can then write the total signal to noise ratio as:
\be
\label{eq:weighted_s2n}
({\rm SN}_{\mathcal{W}})^2 = \sum_\ell \frac{(\ell+1/2)\Delta\ell f_{\rm sky}}{1+(C_\ell^{\rm HI-HI}+N_\ell)\frac{\langle b_{\rm g}{\mathcal{W}}\rangle ^2C_\ell^{\rm DM,b} + \langle {\mathcal{W}} \rangle ^2 C^{\rm shot}}{\langle C_\ell^{\rm HI-\mu} {\mathcal{W}} \rangle ^2}},
\ee
where the average of a quantity $x$ weighted by the number of galaxies $N_{{\rm g},i}$ per magnitude bin $m_i$ is denoted as
$\langle x \rangle = \sum_i x(m_i) N_{{\rm g},i}/\sum_i N_{{\rm g},i}$. We also note that the dark matter power spectrum in the background is given by $C_\ell^{\rm DM,b}(z) = H_0/c \int dz E(z) (W/r)^2 P_{\rm m} ((\ell + 1/2)/r, z)$.

We utilise the weighting function from \citet{Bartelmann:1999yn}, $\mathcal{W} = (5s_{\rm g}-2)/2 $, but we note that in \citet{2011MNRAS.415L..45Y} this weight is generalised to include scale dependence. This can further improve the overall signal to noise ratio (especially for cases where the shot noise is low), but here we use the simplest version that is sufficient for conservative forecasting.

We present the results using this approach in Table~\ref{tab:res_weighted}.
They offer an improvement over the results from subsection~\ref{ssec:original_results} up to a factor of $3$ in the cumulative signal to noise ratio, depending on the foreground redshift. This is expected as this method boosts the signal and uses all available galaxies, keeping the shot noise to a minimum -- we will discuss this further in our conclusions (section~\ref{sec:conclusion}). We now proceed to present forecasts for HI parameters.

\begin{table}
  \begin{center}
    \begin{tabular}{l|cccc}
      &SKA1 B1&-&-&- \\
      $z$ range & 0.34-0.84 & 0.84-1.34 & 1.34-1.84 & 1.84-2.34 \\
      SN$_\mathcal{W}$       &  16.0 &  7.1 & 2.6 &  0.7 \\
      $\frac{\delta(\Omega_{\rm HI}b_{\rm HI})}{(\Omega_{\rm HI}b_{\rm HI})}$      & 0.06  & 0.14  & 0.38  & 1.4 \\ \\
      \hline
       & HIRAX  & - & - & SKA1 B2\\
     $z$ range  & 0.78-1.28 & 1.28-1.78 & 1.78-2.28 & 0.0-0.47\\
     SN$_\mathcal{W}$        &  57.5 & 21.5 &  6.3 &  18.8\\
     $\frac{\delta(\Omega_{\rm HI}b_{\rm HI})}{(\Omega_{\rm HI}b_{\rm HI})}$      & 0.02  & 0.05  & 0.16  &  0.05 
    \end{tabular}
    \captionsetup{width=.9\linewidth}
    \caption{Cumulative signal to noise ratio derived using the weighted 
    galaxy over-density described in section~\ref{ssec:weighted_results}.
    Not requiring a magnitude cut, this method gives best results as all available
    galaxies are used. We also present the forecasted fractional errors on the $\Omega_{\rm HI}b_{\rm HI}$. Note that we fix all other cosmological parameters when calculating these constraints.}
    \label{tab:res_weighted}
  \end{center}
\end{table}

We now discuss what can be learned about the combination of HI abundance and bias, $\Omega_{\rm HI}b_{\rm HI}$, from the expected measurements. These parameters are very important for galaxy evolution studies but remain poorly constrained \citep{Crighton:2015pza}. 
Intensity mapping is a unique and very effective way to provide unprecedented constraints on HI parameters across a wide redshift range. For forecasts using HI clustering in auto and cross-correlation with optical galaxy surveys we refer the reader to \citep{Pourtsidou:2016dzn}.

Arguably, the detections we are presenting here are not expected to be competitive with forthcoming cosmic shear measurements with regards to cosmological parameter constraints. In addition, HI parameters are degenerate with cosmological parameters. Hence, the best use of this data when they become available (at least in the first instance) would be to keep the cosmology fixed and focus on measuring the HI parameters.
This gives the $\Omega_{\rm HI} b_{\rm HI}$ fractional error simply as:
\be
\frac{\delta(\Omega_{\rm HI} b_{\rm HI}(z))}{\Omega_{\rm HI} b_{\rm HI}(z)} 
= 1/{\rm SN}_{\cal W} \, .
\ee
Our derived ${\rm HI}$ constraints are summarised in Table~\ref{tab:res_weighted}. As an example, for our highest signal to noise ratio ${\rm SN}_{\cal W}=51.6$ at $z \simeq 1$ we get a $\sim 2\%$ error, which is much better than currently available measurements at this redshift \citep{Crighton:2015pza}.
We emphasise that these constraints assume all cosmological parameters fixed. This is common in ${\rm HI}$ intensity mapping forecasts, see e.g. \citet{Bacon:2018dui}.
Although the latest constraints on the standard 6 parameter cosmological model are about 1\% for each parameter \citep{Aghanim:2018eyx}, this can still have an impact on the constraints we provided because there are strong degeneracies. This is particularly relevant when constraints on $\Omega_{\rm HI} b_{\rm HI}$ reach below the 10\% level. It will be interesting to account for this by performing a Fisher matrix analysis in future work.

\section{Conclusions}
\label{sec:conclusion}
In this paper we proposed the use of HI intensity maps from large sky surveys with forthcoming radio arrays in cross-correlation with background optical galaxy samples from Stage IV photometric surveys, in order to detect the cosmic magnification signal. 

We then derived predictions for the signal-to-noise ratio of the magnification signal from the foreground HI maps acting on background galaxies using two distinct methods. For both we considered the survey combinations HIRAX with LSST and SKA1-MID with LSST. In \ref{ssec:original_results} the signal-to-noise was optimised by changing the galaxy magnitude threshold $m^*$ for LSST, since a lower magnitude cutoff boosts the magnification signal. 
Due to their different resolutions and mode operations, the information provided by the HIRAX interferometer is complimentary to the data gathered by SKA1-MID in auto-correlation (single dish) mode. 

Then, in subsection \ref{ssec:weighted_results}, we presented a different approach that significantly improves the expected detections in low-redshift bins by using a weighted galaxy over-density.
This method allows to always use all detectable galaxies and thus keeps shot noise at a minimum, whereas the method described in \ref{ssec:original_results} requires magnitude cutoffs in the galaxy samples. The lower the foreground redshift, the more stringent these optimised cuts and thus the difference between both approaches increases. When the galaxy samples in the weighted summation are restricted to the same magnitude cuts as in \ref{ssec:original_results}, the forecasts of both methods agree better. 

A detection seems likely with forecasted cumulative signal to noise ratios in the range of $\sim 50$, but a more detailed analysis with appropriate simulations  will be needed to fully assess all relevant sources of errors, e.g. foreground contamination residuals and cleaning effects.
Foreground residuals are not expected to be significant in the cross-correlation between HI intensity maps and galaxies. The loss of long-wavelength radial modes in the HI data is also not expected to have a significant deteriorating effect on this observable.
Foreground contamination is expected to be most severe on large angular scales, as shown in e.g. \citet{Wolz:2013wna,Cunnington:2019lvb}. We assessed the sensitivity of our analysis to this by entirely discarding the lowest three $\ell$ bins (i.e. $\ell <40$ for SKA and $\ell < 240$ for HIRAX). Even with the very pessimistic assumption that no information can be obtained from these bins, the signal to noise ratio is only reduced by $\sim 3$\% and $\sim 6$\% for HIRAX and SKA, respectively.
However, it would be useful to properly account and quantify all foreground effects by extending the cross-correlation simulations studies performed in \citet{Witzemann:2018cdx, Cunnington:2019lvb,Cunnington:2018zxg} -- we leave this for future work.
We also note that the choice of redshift binning could be reconsidered to make the analysis more realistic for a foreground cleaned HI survey. Furthermore, using realistic simulated LSST catalogues we can implement and test the performance of scale-dependent optimal weighting functions \citep{2011MNRAS.415L..45Y}.

Finally, we turned the expected signal to noise ratio into a fractional constraint on $\Omega_{\rm HI}b_{\rm HI}(z)$. These HI parameters are still poorly constrained, but we show that a cosmic magnification analysis can yield a fractional error as good as $\sim 2$\%, if cosmological parameters are fixed.

To conclude, both methods presented in this work give results suggesting that it will certainly be possible to detect the magnification signal once the data is available. This will be complementary to measurements using optical foreground samples with completely different systematics.

\section*{Acknowledgments}
We are extremely grateful to the anonymous referee, whose suggestions greatly improved the quality and scope of this manuscript. We thank David Bacon for useful discussions. 
AW and MGS acknowledge support from the South African Square Kilometre Array Project and National Research Foundation (Grant No. 84156). 
AP is a UK Research and Innovation Future Leaders Fellow, grant MR/S016066/1, and also acknowledges support from the UK Science \& Technology Facilities Council through grant ST/S000437/1.

\bibliographystyle{mnras}
\bibliography{magnification_IM}

\end{document}